\documentclass[sigplan,screen]{acmart}
\acmConference[PPoPP'25]{ACM SIGPLAN Conference on Programming Languages}{March 01--05, 2025}{Las Vegas, NV, USA}
\acmYear{2025}
\acmISBN{} 
\acmDOI{} 
\startPage{1}

\setcopyright{none}

\usepackage{booktabs}   
\usepackage{subcaption} 
\usepackage{algorithm}
\usepackage{algpseudocode}
\usepackage{makecell}

\usepackage{natbib}
\usepackage{etex}

\usepackage{graphics}
\usepackage{epsfig}
\usepackage{svg}

\usepackage{pifont}
\usepackage{enumitem}
\usepackage{svg}
\usepackage{enumitem}
\usepackage{lipsum}
\usepackage[normalem]{ulem}
\setitemize[1]{itemsep=0pt,partopsep=0pt,parsep=\parskip,topsep=5pt}
\setdescription{itemsep=0pt,partopsep=0pt,parsep=\parskip,topsep=5pt}
\usepackage{titlesec}

\author{
    Haisha Zhao$^{1,2,}$\footnotemark[1],
    San Li$^{1,2,}$\footnotemark[1],
    Jiaheng Wang$^{3}$,
    Chunbao Zhou$^{1,2}$,
    Jue Wang$^{1,2,}$\footnotemark[2],
    Zhikuang Xin$^{1,2}$,
    Shunde Li$^{1,2}$,
    Zhiqiang Liang$^{1,2}$,
    Zhijie Pan$^{4}$,
    Fang Liu$^{1,2}$,
    Yan Zeng$^{4}$,
    Yangang Wang$^{1,2}$,
    Xuebin Chi$^{1,2}$
}

\affiliation{
    \institution{$^{1}$Computer Network Information Center, Chinese Academy of Sciences, Beijing, China}
    \institution{$^{2}$University of Chinese Academy of Sciences, Beijing, China}
    \institution{$^{3}$Renmin University of China, Beijing, China}
    \institution{$^{4}$Hangzhou Dianzi University, Hangzhou, Zhejiang, China}
    \country{}
}

\email{hszhao@cnic.cn, lisan24@mails.ucas.ac.cn, wangjiaheng0413@ruc.edu.cn}
\email{zhoucb@cnic.cn, wangjue@sccas.cn, xinzhikuang@cnic.cn, lishunde@cnic.cn, zqliang@cnic.cn}
\email{panzhijie@hdu.edu.cn, liufang@sccas.cn, yz@hdu.edu.cn, wangyg@sccas.cn, chi@sccas.cn}

\begin{document}
\pagestyle{plain}
\title[Acc-SpMM: Accelerating General-purpose SpMM with GPU Tensor Cores]{Acc-SpMM: Accelerating General-purpose Sparse Matrix-Matrix Multiplication with GPU Tensor Cores}         





\renewcommand{\shortauthors}{Zhao, Li and Wang et al.}
\renewcommand{\thefootnote}{\fnsymbol{footnote}}

\begin{abstract}
\thispagestyle{plain}

General-purpose Sparse Matrix-Matrix Multiplication (SpMM) is a fundamental kernel in scientific computing and deep learning. The emergence of new matrix computation units such as Tensor Cores (TCs) brings more opportunities for SpMM acceleration. However, in order to fully unleash the power of hardware performance, systematic optimization is required. In this paper, we propose Acc-SpMM, a high-performance SpMM library on TCs, with multiple optimizations, including data-affinity-based reordering, memory efficient compressed format, high-throughput pipeline, and adaptive sparsity-aware load balancing. 
In contrast to the state-of-the-art SpMM kernels on various NVIDIA GPU architectures with a diverse range of benchmark matrices, Acc-SpMM achieves significant performance improvements, on average 2.52x (up to 5.11x) speedup on RTX 4090, on average 1.91x (up to 4.68x) speedup on A800, and on average 1.58x (up to 3.60x) speedup on H100 over cuSPARSE. 
\end{abstract}


\maketitle
\footnotetext[1]{ Joint first authors.}
\footnotetext[2]{ Corresponding author.}

\vspace{-2mm}
\section{Introduction}
General-purpose Sparse Matrix-Matrix Multiplication (SpMM) is a fundamental and expensive computational kernel in numerous scientific computing, data mining and deep learning, such as linear algebra solvers \cite{Vzquez2010AMA, Aktulga2014OptimizingSM, Kepner2016MathematicalFO, Demmel2000TemplatesFT}, linear algebra calculations \cite{Kannan2015AHP, Moon2020ALONMFAL, Ahmad2023ExploringDL}, graph analysis \cite{langville2006google, si2014multi, Yang2019GraphBLASTAH, Dai2004AdvancesIK}, Graph Neural Networks (GNNs) \cite{Peng2023MaxKGNNEF, Gao2023AlgorithmHardwareCF, Yin2023DGIAE, HoqueAnik2024iSpLibAL} and large-scale deep learning models \cite{NEURIPS2020_1457c0d6, Hoefler2021SparsityID, Child2019GeneratingLS}. 
Therefore, optimization of SpMM has the potential to impact a wide variety of applications.

Extensive related studies have been conducted to improve the performance of SpMM on the general-purpose computing units of GPUs, 
utilizing a variety of optimization techniques \cite{Dai2022HeuristicAT, Gale2020SparseGK, Hong2019AdaptiveST, Huang2020GESpMMGS, Jiang2020AND}, 
including sparse storage formats, reordering algorithms, parallel strategies, and memory access optimizations.
With the emergence of specialized GPU hardware, particularly Tensor Cores (TCs), there are more promising opportunities to accelerate SpMM computations \cite{Fan2024DTCSpMMBT, Li2022EfficientQS, Chen2021EfficientTC, Castro2022ProbingTE}.

Nevertheless, there are still obstacles to fully unleashing the potential of TCs to achieve a substantial acceleration of SpMM. 
\ding{182} SpMM is classified as a memory access-intensive computation due to its irregular memory access patterns.
The sparse storage format impacts not only the memory footprint but also the efficiency of memory access.
Researches show that sparse matrix format innovation is effective to improve the performance of SpMM \cite{Lpez2014FastSpMMAE, nvidia2021, Anzt2015AcceleratingTL, Zhang2018RegularizingIB}. 
Existing sparse storage formats either have low compression efficiency or high overhead.
\ding{183} TCs are designed to operate on dense data, 
which may not be a natural match for sparse data operations in SpMM. 
Meanwhile, tiling is a common technique in the implementation of reordering algorithms \cite{Arai2016RabbitOJ, Jiang2020AND, Hong2019AdaptiveST, Kurt2020EfficientTS}. 
Data locality improves as the number of non-zero elements (nnz) within a tiling block increases, which can lead to enhanced computational performance for SpMM.
Balancing data locality with reordering overhead is a challenge for current reordering algorithms. 
\ding{184} Instruction-Level Parallelism (ILP) is typically employed to overlap memory access and computation on GPUs, a technique known as pipeline, which relies on the arrangement of independent memory and computation operations \cite{Fan2024DTCSpMMBT, Yan2020DemystifyingTC, Crago2024WASPEG, Crago2018ExposingMA}. 
To the best of current knowledge, existing memory access optimization methods are relatively simple and inefficient, 
often characterized by low memory bandwidth or lots of pipeline bubbles.
 
In order to address the aforementioned challenges, we propose Acc-SpMM, a novel approach that is specifically tailored to TCs and incorporates systematic techniques and optimizations.

In this paper, the main contributions are given as follows.



\begin{itemize}

\item We propose a novel sparse matrix compression storage format that is memory-efficient and incurs little overhead for decompression.
\item We design a data-affinity-based reordering algorithm with $O(nlogn)$ complexity that efficiently handles fine-grained tiling data while satisfying data locality requirements.
\item We build a highly optimized SpMM kernel, which incorporates an advanced pipeline method with multi-level memory access techniques. Our proposed SpMM kernel decreases the number of bubbles and overlaps computations with multiple memory access, enhancing memory and compute throughput.
\item We propose an adaptive sparsity-aware load balancing method based on an effective performance model with writing back cost involved, which highly increases both memory throughput and compute throughput.
\item We develop Acc-SpMM that accelerates general SpMM, which achieves up to 5.11x speedup on RTX 4090, up to 4.68x speedup on A800, and up to 3.60x speedup on H100, compared to state-of-the-art TC-based SpMM approaches.
\end{itemize}

We evaluate Acc-SpMM on large-scale power-law graph matrices in GNNs and sparse matrices from the SuiteSparse Matrix Collection \cite{Davis2011TheUO} extensively. 
We also compare the efficacy of the proposed Acc-SpMM with TCGNN-SpMM \cite{wang2023tc}, Sputnik\cite{Gale2020SparseGK}, SparseTIR\cite{Ye2022SparseTIRCA}, DTC-SpMM \cite{Fan2024DTCSpMMBT} and the widely-used SpMM kernel in the cuSPARSE library \cite{Naumov2010} on recent RTX4090 (Ada Lovelace) \cite{nvidia_rtx_4090_architecture}, A800 (Ampere) \cite{nvidia_a100} and H100 (Hopper) \cite{H100_Tensor_Core_GPU_Architecture} GPUs. 
Experimental results demonstrate that Acc-SpMM achieves noteworthy average speedups over alternative approaches on diverse real-world matrices. 


\section{Background and Motivation}
\subsection{NVIDIA GPU and Tensor Core}

NVIDIA GPUs are highly parallel processor architectures that feature a memory hierarchy and various processing elements, including multiple Streaming Multiprocessors (SMs), an on-chip L2 cache, and high-bandwidth DRAM. 
All SMs share access to the L2 cache and DRAM.
Each SM has a set of Streaming Processors (SPs) which has its own register file, scheduler and execution units. 
All SPs within the same SM share a private L1 cache, part of which can be configured as the shared memory. 
Each SP has Floating-point Units (FPUs) and Tensor-core Units (TCUs). 
The FPU is the traditional CUDA core that can perform fused multiply-and-accumulate (FMA) operations per cycle. 
The TCU can perform one Matrix Multiply-and-Accumulate (MMA) operation per cycle, delivering higher performance than FPU on CUDA core.
Kernels are executed by SMs with 2-level hierarchy of threads. 
In SIMT mode, threads within a warp (32 threads) execute the same instruction simultaneously \cite{nvidiaCUDAToolkit}.

TCs are specialized cores that enable mixed precision training since the Volta architecture \cite{NVIDIA_Volta_GPU_Architecture_Whitepaper}. 
With each subsequent generation, more datatypes are involved for computation with the new GPU architectures, including TF32 and FP64 \cite{nvidia_a100, H100_Tensor_Core_GPU_Architecture}. 
On Volta and Turing, there are $8$ TCUs on each SM, whereas on Ampere, AdaLovelace, and Hopper, the number of TCU per SM decreases to $4$. 
Despite the reduction in the number of TCUs on each SM since the Ampere architecture, the throughput per cycle has increased by $4$ times. 
Consequently, the overall performance of TCUs per SM has been doubled.


CUDA provides warp-level APIs with MMA semantics. 
Warp-level Matrix Multiply Accumulate (WMMA) \cite{nvidiaCUDAToolkit} in C++ performs a dense matrix multiplication with contiguous memory constraints. 
The PTX-level MMA \cite{nvidia_ptx_isa} can perform matrix multiplications with non-contiguous memory, providing more possibilities for sparse matrix multiplication. 
Both of them are warp level matrix multiplication. The shape of m16n16k8 api multiplies a left-handed matrix of shape $(16 \times 8)$ with a right-handed matrix of shape $(8 \times 16)$. 

\subsection{Motivation of SpMM on Tensor Cores}

\noindent\textbf{High memory consumption}: 
A matrix of shape $m \times n$ is generally referred to as sparse if the number of its nnzs is small enough compared to $O(mn)$. 
Compressed storage formats are used in SpMM, such as Compressed Sparse Row (CSR) \cite{Saad2003} and Coordinate (COO) \cite{templates} format, to make it possible to process sparse matrices with very large scales. SpMM is a memory-bound kernel for the irregular data access pattern brought by sparse data structures. 
For this reason, the memory access efficiency of dense matrices may lead to substantially varying performance across a variety of compressed storage formats.

\noindent\textbf{Low density and low locality of matrix}: 
The nnzs in a sparse matrix are always stored in a row-major pattern, such as CSR and COO formats. 
However, because the nnzs within the same row can have a wide range of column indices, the memory access pattern for the corresponding dense matrix becomes non-contiguous.
Additionally, sparse matrices must be divided into smaller blocks, referred to as TC blocks in our paper, to accommodate the features of the TCUs during matrix multiplication operations.
Hence, the computational performance of SpMM is proportional to both the density of TC blocks (the TCU utilization) and the data locality of TC blocks (the memory access efficiency).

\noindent\textbf{Low pipeline utilization}: 
The efficiency of TCs depends on the utilization of the TC pipeline. 
Low utilization is primarily caused by two factors.
Firstly, accessing global memory is inefficient. 
SpMM involves loading a sparse matrix $A$, a dense matrix $B$, and storing a dense matrix $C$. 
The cache hit rate may decrease because all memory accesses from global memory are by default passed through the L1 and L2 caches, leading to frequent cache misses.
Secondly, the overhead associated with data access requests is high. 
The TCUs remain idle during memory access operations by default. 
Although several buffering methods have been used to overlap computation and memory access, there is still opportunities for further improvement in pipeline utilization.
\section{Acc-SpMM}

\subsection{Overview}

\begin{figure}[h]
    \centering
    \includegraphics[width=0.9\linewidth]{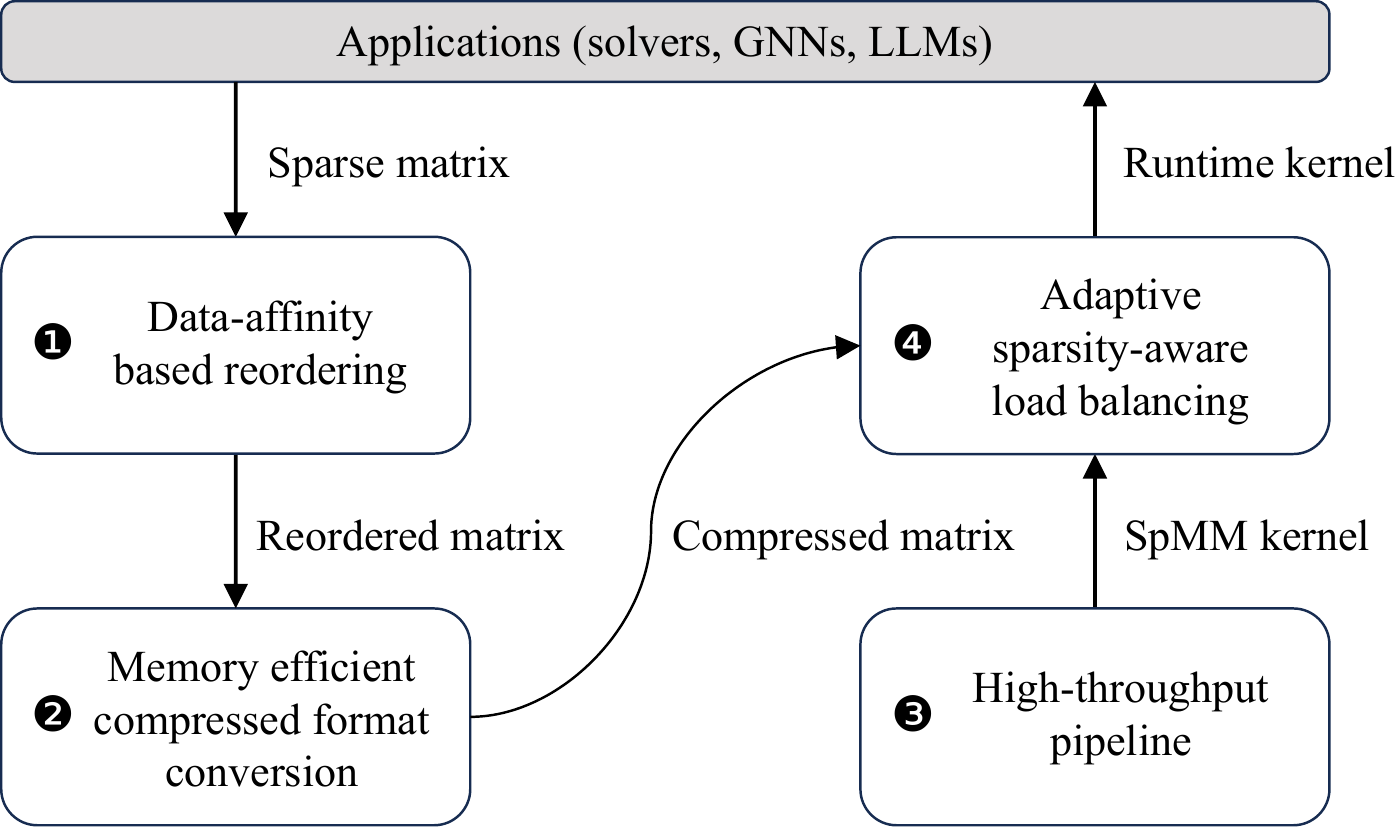}
    \caption{The overview of Acc-SpMM}
    \label{figOverview}
\end{figure}

Acc-SpMM consists of four key components, as shown in Figure \ref{figOverview}:
\ding{182} The data-affinity-based reordering splits and rearranges the sparse matrix into matrix blocks to enhance the density of TC blocks and improve data locality. 
\ding{183} The memory-efficient compressed format translation converts sparse matrix blocks into our proposed BitTCF format to improve memory access efficiency. 
\ding{184} The high-throughput pipeline executes an efficient runtime kernel.
With efficient memory access, the high-throughput pipeline is fully utilized to overlap memory access, thereby improving TC pipeline utilization.
\ding{185} Adaptive sparsity-aware load balancing dynamically adjusts the number of TC blocks assigned to each thread block (TB) based on the sparsity of the matrix.

\subsection{Data-affinity-based reordering}

TC is naturally suitable for the calculation of dense matrices, therefore, ensuring data density within each TC block is crucial for achieving optimal TCU computational efficiency. 
Simultaneously, improving data locality is critical for enhancing the utilization of hierarchical caches on GPUs.
In light of above and inspired by modularity-based reordering algorithm \cite{Arai2016RabbitOJ}, we propose a novel reordering algorithm to achieve high data density and locality.

Modularity \cite{newman2004finding}, which is widely used to evaluate the quality of the community structure in a graph.
$\Delta Q$ is defined in Equation (\ref{equa-modulrity}), where $\Delta Q$ is the modularity improvement, $m$ is the total edges in the graph, $A_{ij}$ is the weight between node $i$ and node $j$, $k_i$ is the degree of node $i$, $s_i$ is the community that involves node $i$ and $\delta_{s_i,s_j}$ is the Kronecker delta function, which equals 1 when $s_i = s_j$ and 0 otherwise. 
The graph is constructed by using a sparse matrix as the adjacency matrix, where each node in the graph corresponds to an index of a row or a column.
If there is a nnz in the matrix, the weight between the corresponding nodes is typically set to 1, otherwise, it is set to 0. After dendrogram construction, the graph is divided into different communities. 
By minimizing the number of cross-community edges and arranging vertexes with similar edges closer, data-affinity-based-reordering effectively makes the arrangement of nnzs more compact locally.

\begin{figure}[thpb]
    \centering
    \includegraphics[width=\linewidth]{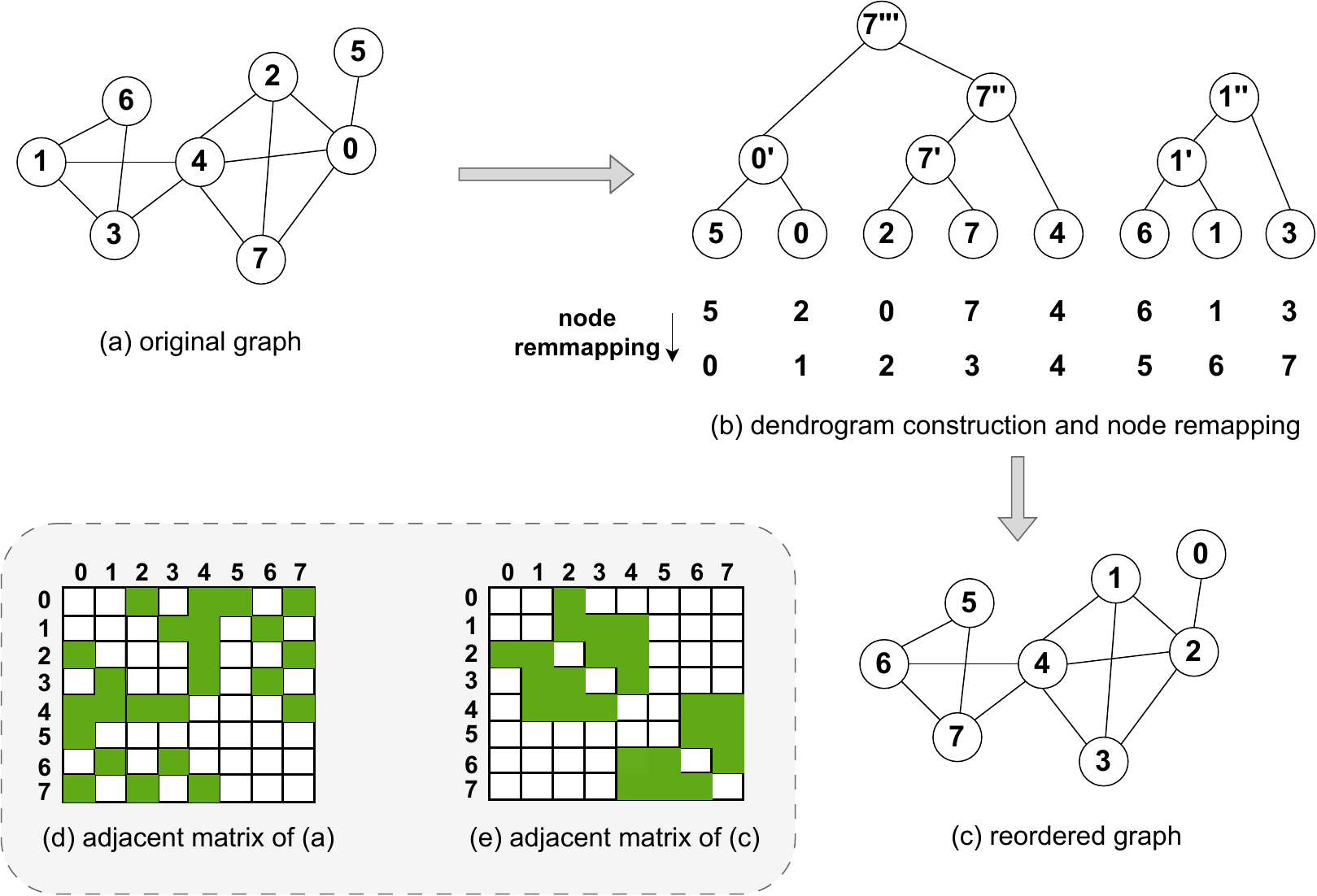}
    \caption{The design of data-affinity-based reordering. (a) is the original graph, (b) is dendrogram construction and node remapping, (c) is the reordered graph, (d) is the adjacency matrix of the original graph (a), and (e) is the adjacency matrix of the reordered graph (c).}
    \label{order-graph}
\end{figure}

Algorithm \ref{reorder} outlines data-affinity-based reordering. 
Given a sparse matrix, the algorithm outputs a reordered sparse. 
Firstly, we generates the dendrogram from a sparse matrix in the graph structure by coarsening and refinement (line 1-8). Then, we apply ordering generation (line 9-27), which converts the dendrogram into a new ordered graph.

\textbf{Dendrogram construction}. 
Dendrogram construction can improve the density for each community. We firstly pick a source vertex $v$ in ascending order of degree (line 2), then find the neighbor vertex $u$ of $v$ that gains the maximum $\Delta Q$ (line 4), and finally merge the source vertex $v$ with the neighbor vertex $u$ if the modularity improves (lines 5–7).
\begin{displaymath}
\small
\label{equa-modulrity}
\Delta Q = \frac{1}{2m} \sum_{i,j}\left(A_{ij} - \frac{k_ik_j}{2m}\right) \delta_{s_i,s_j}
\tag{1}
\end{displaymath}
\vspace{1mm}

\textbf{Ordering generation.} 
Ordering generation can improve data locality for each TC block and optimize the utilization of hierarchical caches on GPUs.
After constructing the dendrogram, a depth-first search (DFS) is performed, and the first visited leaf node is selected as the source vertex $v$ (line 11). The ordering for vertex $v$ is then generated (line 15), followed by selecting a vertex $u$ from the leaf nodes that shares the maximum number of common neighbors with $v$ (line 18). Then, the ordering for vertex $u$ is generated (line 22), and $u$ is set as the new source vertex (line 25). This process continues iteratively until all vertices are visited.

\begin{algorithm}
    \caption{Data-affinity-based reordering}
    \renewcommand{\algorithmicrequire}{\textbf{Input:}}
    \renewcommand{\algorithmicensure}{\textbf{Output:}}
    \label{reorder}
    \begin{algorithmic}[1]
    \small
    \Require Sparse Matrix $A$
    \Ensure Sparse Matrix $A_{reordered}$
    \State{\textcolor{blue}{$\triangleright$ Step I: dendrogram construction}} 
    \State{$V \gets$ vertex list of $A$}
    \For{$v \in V$ in ascending order of degree}
    \State{$u \gets $ nbr of $v$ which maximize $\Delta Q(u,v)$}
    \If{$\Delta Q(u,v) > 0$}
    \State{merge $v$ into $u$ and record this merge in $dendrogram$}
    \EndIf
    \EndFor
    \State{\textcolor{blue}{$\triangleright$ Step II: ordering generation}} 
    \State{$new\_vid \gets 0$}
    \For{$v \in V$ in DFS on dendrogram}
    \If{$v$ is visited}
    \State{continue}
    \EndIf
    \State{$v \gets new\_vid$}
    \State{$new\_vid \gets new\_vid + 1$}
    \State{mark $v$ as visited}
    \While{$u \in V$ in DFS that has most common nbrs with $v$}
    \If{$u$ is visited}
    \State{continue}
    \EndIf
    \State{$u \gets new\_vid$}
    \State{$new\_vid \gets new\_vid + 1$}
    \State{mark $u$ as visited}
    \State{$v \gets u$}
    \EndWhile
    \EndFor
    \State{\Return $A_{reordered}$}
    \end{algorithmic}
\end{algorithm}

The diagram of data-affinity-based reordering is shown in Figure \ref{order-graph}. 
After constructing the dendrogram for the original graph in Figure \ref{order-graph} (a), we obtain two communities in Figure \ref{order-graph} (b). 
Then we perform DFS-like search, pick a source vertex 5 of the first community, and set its new order index to 0. We search all the leaf nodes in the dendrogram according to DFS that have not been visited and try to find the vertex which has the maximal number of common neighbors with source vertex 5. Vertex 2, 7, and 4 all have a common neighbor vertex 0 with vertex 5, and we choose vertex 2 as the next visited node according to the order of DFS. 
Subsequently, we assign an order of 1 to vertex 2 and designate vertex 2 as the source vertex. 
We then continue to perform DFS until all the nodes in the dendrogram are visited.
After all vertexes have been reordered, we obtain the reordered graph in Figure \ref{order-graph} (c). 
We apply SpMM according to the reordered graph. 
Compared to adjacency matrix in Figure \ref{order-graph} (d) for original graph in Figure \ref{order-graph} (a), we get better data density and locality in Figure \ref{order-graph} (e) for reordered graph in Figure \ref{order-graph} (c) through reordering.

\subsection{Memory efficient compressed format}

\begin{figure}[thpb]
    \centering
    \includegraphics[width=\linewidth]{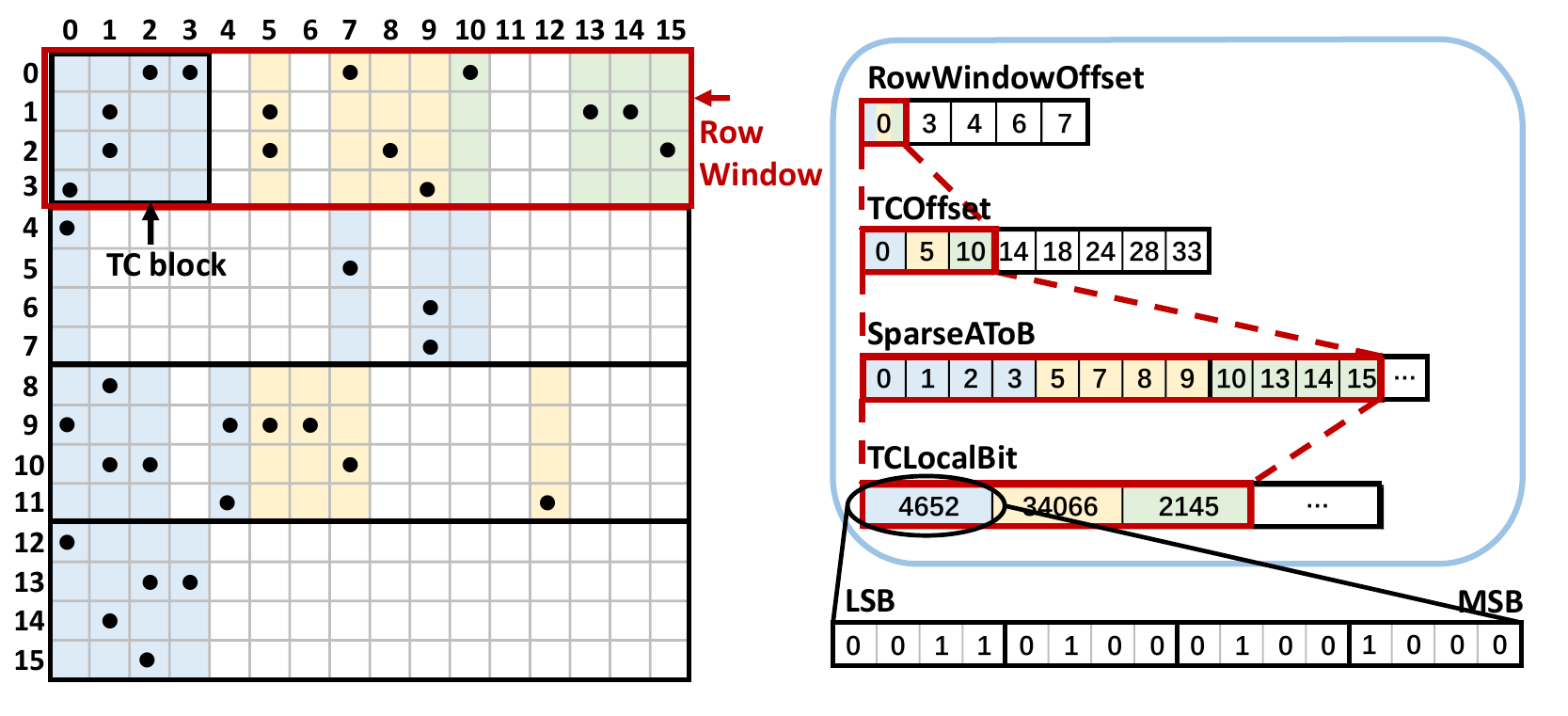}
    \caption{The design of BitTCF format.}
    \label{figBME}
\end{figure}

We propose a memory efficient compressed format based on ME-TCF \cite{Fan2024DTCSpMMBT}, which we call BitTCF as shown in Figure \ref{figBME}. For convenience in illustration, a $4 \times 4$ tile is shown in Figure \ref{figBME} to demonstrate the partitioning of the sparse matrix $A$. However, we choose the shape of $8 \times 8$ tile in reality. BitTCF utilizes four arrays to represent a sparse matrix. \ding{182} $RowWindowOffset$ represents the offset of the starting TC block in each RowWindow. 
In the case of $8\times 8$ TC blocks, $RowWindowOffset$ contains $\lceil M/8 \rceil + 1$ elements. 
\ding{183} $TCOffset$ holds the offset of the starting nnz in each TC block, which contains $NumTcBlock + 1$ elements. 
\ding{184} $SparseAToB$ holds the original column indices of each nnz in TC blocks and contains $ NumTcBlock \times 8$ elements. \ding{185} We use a \textbf{uint64} integer to represent the local position of each nnz in each TC block, which we call $TCLocalBit$, where $1$ indicates a nnz and $0$ denotes a zero element. In total, BitTCF requires $(\lceil M/8 \rceil + NumTCBlock \times 11 + 2) \times 4$ bytes to represent a sparse matrix, where $M$ is the number of rows of sparse matrix $A$, and $NumTCBlock$ is the number of TC blocks in matrix $A$.

The size of $TCLocalId$ in ME-TCF corresponds to the number of nnzs, representing original positions of each nnz.
$TCLocalId$ stores all the positions of nnzs in each TC block with a \textbf{int8}, and in a $8 \times 8$ tile of TC block, there are at least 8 nnzs and at most 64 nnzs, which will cost at least 64 bit and at most 512 bit to store the position of each nnz in a TC block with ME-TCF format. We design $TCLocalBit$ to utilize a uint64 integer to store all the positions of the nnzs in a TC block. 
For each TC block, we only need 64 bit to store all the positions of nnzs. 
To better use uint64 for compression, we partition the sparse matrix A into $8\times8$ tiles.
Compared to ME-TCF, BitTCF can effectively save memory as the number of nnzs increases.
Meanwhile, in order to take advantage of the multi-level memory and cache structure of the GPU, we design a memory access pattern for BitTCF. 
In this pattern, $SparseAToB$ is loaded from global memory to shared memory for reuse, while $RowWindowOffset$, $TCOffset$ and $TCLocalBit$ are loaded from global memory to registers directly.
This approach does not impact the cache hit rate for the frequently used dense matrix.
The compressed format and memory access pattern can improve memory access efficiency. 

Additionally, we leverage C++ bit operations to deliver decompression. During the decompression phase, we use bit-wise operations to decompress the offsets of nnz positions. We use two warps for decoding, with thread IDs ranging from 0 to 63. Each thread is responsible for decompressing the element at a specific position. Each thread checks whether the current position is a nnz. If it does, the \textbf{\_\_popcll} api is used to calculate the offset of the nnz at that position. Otherwise, it directly writes zero into shared memory. The overhead of the decompression is minimal. Moreover, decompression can be overlapped with memory access.

\subsection{High-throughput pipeline}

Algorithm \ref{pipeline-algorithm} outlines a runtime kernel that integrates a high-throughput least bubble double-buffers pipeline. 
Given a sparse matrix $A$, a $SparseAToB$, and a dense matrix $B$, the algorithm outputs the dense matrix $C$ after applying SpMM. The runtime kernel arranges the memory locations (shared memory or registers) of multiple arrays to maximize cache hit rate. The double-buffers pipeline arranges the memory access pattern of multiple arrays to maximize TCU utilization.

\begin{figure}[thpb]
    \centering
    \includegraphics[width=0.9\linewidth]{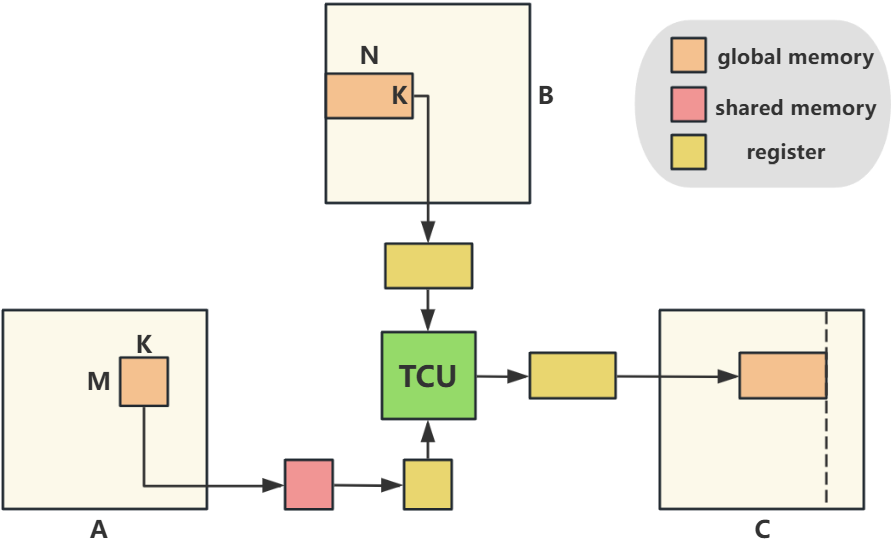}
    \caption{The data movement of Acc-SpMM. TC block is moved to shared memory for reuse. Dense $B$ tile is moved directly to registers. Dense matrix $C$ is moved to global memory from registers.}
    \label{mma}
\end{figure}

\textbf{Runtime kernel}. The data movement of the runtime kernel is shown in Figure \ref{mma}. Runtime kernel loads sparse matrix $A$ tiles from global memory to shared memory (line 21), loads array $SparseAToB$ from global memory to shared memory (line 24), loads dense matrix $B$ tiles from global memory to registers directly (line 21), and stores dense matrix $C$ tiles from registers to global memory (line 37) after computation for each RowWindow of sparse matrix A. 
As shown in Table \ref{cache-control}, we can use PTX-level instructions to control the cache policy to improve the cache hit rate. Due to the limit of asynchronous copy instruction, we cache sparse matrix $A$ in both L1 and L2 caches using the \textbf{.ca} instruction.
Since we need to access dense matrix $B$ multiple times, we cache it in both L1 and L2 caches to enhance the cache hit rate for dense $B$ matrix using the \textbf{.ca} instruction. We do not need to reload dense matrix $C$ again after mma computation, so we choose not to cache result dense matrix $C$ in L2 cache, using the \textbf{.wt} instruction. 


\begin{table}[thpb]
\small
\caption{Cache operators for memory instructions.}
\label{cache-control}
\begin{tabular}{cc}
\hline
operator             & Meaning\\
\hline
.ca                  & Cache at all levels, likely to be accessed again\\
.cg                  & Cache in L2 and below, not L1\\
.cs                  & Cache streaming, likely to be accessed once \\
.lu                  & Last use\\
.cv                  & Don't cache and fetch again\\
\hline
.wb                  & Cache write-back all coherent levels\\
.wt                  & Cache write-through the L2 Cache\\
\hline
\end{tabular} \\
\end{table}



In this paper, we mainly focus on \textbf{tf32} datatype, which is the most commonly used datatype in GNNs. Only the \textbf{m16n8k8} and \textbf{m16n8k4} shapes of the mma api support tf32 calculation on TCs\cite{nvidia_ptx_isa}. We choose \textbf{m16n8k8} due to its lower synchronization cost.
When applying \textbf{m16n8k8}, a shape of $16 \times 8$ matrix tile and a shape of $8 \times 8$ matrix tile are required. 
We load the matrix followed by swapping the computation of the left-handed matrix and the right-handed matrix (lines 28 and 35). This swapping enables us to partition the sparse matrix $A$ into $8 \times 8$ TC blocks (line 26), which significantly improves the density of TC blocks\cite{shi2024flashsparse}. This arrangement also makes it more advantageous for us to use the uint64 datatype in our compression format. 


\begin{figure}[thpb]
    \centering
    \includegraphics[width=\linewidth]{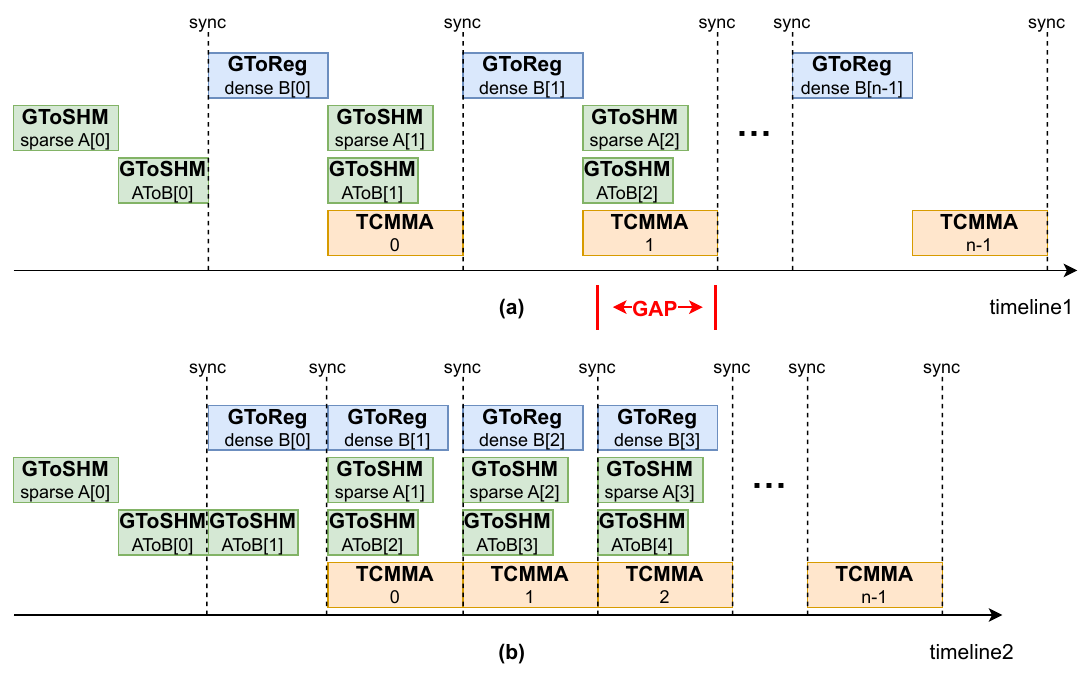}
    \caption{Comparisons between our proposed pipeline(b) with DTC-pipeline(a). GToReg: global memory to register; GToSHM: global memory to shared memory; TCMMA: tensor core mma.}
    \label{figpipeline}
\end{figure}

\textbf{Least bubble double-buffers pipeline}. 
To reduce pipeline bubbles, we propose a high-throughput pipeline, as shown in Figure \ref{figpipeline} (b), which overlaps various memory accesses and TCU MMA operations, thereby further enhancing TC pipeline utilization.

For DTC-pipeline shown in Figure \ref{figpipeline} (a), since mma computation depends on dense matrix $B$, there is an implicit synchronization after GToReg of dense matrix $B$. As a result, the TCUs remain idle while the tiles of dense matrix $B$ are being loaded, leading to significant pipeline bubbles. This prevents full utilization of the memory bandwidth and results in low TCU pipeline utilization. Our proposed pipeline method prefetches the dense matrix $B$ tiles required for the next mma computation, further overlapping data loading with computation to reducing pipeline bubbles. To apply the high-throughput pipeline, we design double buffers in shared memory to store the sparse matrix $A$ tiles, and the $SparseAToB$ arrays. We also use \textbf{cp.async} to handle computation and data loading asynchronously, maximizing the overlap.
In Figure \ref{figpipeline} the GAP shows the execution time difference between our proposed pipeline and DTC pipeline after two TCMMA operations. 
Our proposed pipeline significantly improves TCU pipeline utilization.

\begin{algorithm}
    \caption{Pipeline runtime kernel}
    \renewcommand{\algorithmicrequire}{\textbf{Input:}}
    \renewcommand{\algorithmicensure}{\textbf{Output:}}
    \label{pipeline-algorithm}
    \begin{algorithmic}[1]
    \small
    \Require SpMatrix $A$, DnMatrix $B$, SparseAToB $AToB$
    \Ensure DnMatrix C
    \State $sTCOffset \gets$ RowWindowOffset[blockIdx.x];
    \State $eTCOffset \gets$ RowWindowOffset[blockIdx.x+1]; \State \textcolor{blue}{$\triangleright$ TC block buffers in SHM.}
    \State \_\_shared\_\_ ATile[2][128];    
    \State \textcolor{blue}{$\triangleright$ AToB buffers in SHM.}
    \State \_\_shared\_\_ AToBTile[2][8];      
    \State{register fragB[2];}
    \State \textcolor{blue}{$\triangleright$ Prefetch ATile and AToBTile to SHEM}
    \State $ATile[0] \gets sTCOffset$ in $8 \times 8$ shape;
    \State $AToBTile[0] \gets$ values correspond to $sTCOffset$ in $AToB$;
    \State $synchronize()$;
    \State $AToBTile[1] \gets$ values correspond to $sTCOffset+1$ in $AToB$;
    \State{\textcolor{blue}{$\triangleright$ Load dense $B$ tile for current MMA}} 
    \State{$fragB[0] \gets BTile$ to register in $8 \times 16$ shape remapping.}
    \For{$i \gets sTCOffset + 1$ \textbf{to} $eTCOffset$ \textbf{step} 1}
        \State{\textcolor{blue}{$\triangleright$ select which buffer to read.}}
        \State{$rdSel \gets (i - sTCOffset + 1)\ and\ 1$}
        \State{\textcolor{blue}{$\triangleright$ select which buffer to store.}}
        \State{$stSel \gets (i - sTCOffset)\ and\ 1$}
        \State{\textcolor{blue}{$\triangleright$ Load dense $B$ tile for next MMA}} 
        \State {$fragB[rdSel] \gets BTile$ to register in $8 \times 16$ shape;}
        \State \textcolor{blue}{$\triangleright$ Async copy TC block and AToBTile to SHM.} 
        \State $ATile[stSel] \gets i$-$th$ TC\ block;
        \State $AToBTile[stSel] \gets (i+1)$-$th\ AToB$;
        \State \textcolor{blue}{$\triangleright$ Fetch TC block from SHM to register.}
        \State $fragA \gets ATile[rdSel]$ in $8 \times 8$ shape;
        \State \textcolor{blue}{$\triangleright$ TC computation with swapped mma.}
        \State TCMMA($fragB[rdSel]$, $fragA$, $fragC$); \State \textcolor{blue}{$\triangleright$ Async transaction barrier.}
        \State $WaitGroup();$ 
        \State $synchronize()$;
    \EndFor
    \State{\textcolor{blue}{$\triangleright$ select which buffer to read.}}
    \State{$rdSel \gets (eTCOffset - sTCOffset + 1)\ and\ 1$}
    \State {TCMMA($fragB[rdSel]$, $fragA$, $fragC$);}
    \State \textcolor{blue}{$\triangleright$ Write through $C$ back.}
    \State {WTCREMAPPING($C$, $fragC$); 
    }
    \end{algorithmic}
\end{algorithm}

The detailed pipeline is described in Algorithm \ref{pipeline-algorithm}. In the warm-up phase of pipeline, a sparse matrix $A$ tile and an array $SparseAToB$ are loaded (lines 9 and 10). When an array $SparseAToB$ is already in shared memory, a corresponding dense matrix $B$ tile is loaded (line 14), and another array $SparseAToB$ is loaded at the same time (line 12). In stable phase of pipeline, when applying TCMMA (line 28), a dense matrix $B$ tile, a sparse matrix $A$ tile, and an array $SparseAToB$ are loaded asynchronously (line 21-24). 
The proposed pipeline effectively overlaps the memory access (dense matrix $B$ tiles) with memory access (sparse matrix $A$ tiles and $SparseAToB$ arrays), as well as the memory access with computation. 
In the frozen phase of pipeline, the last TC block of the RowWindow is calculated (line 35), and multiplication results are written to global memory (line 37). For the sake of illustration, the TCMMA duration is intentionally exaggerated for clarity in the diagram. In reality, the computation time is much shorter than the data loading time.

\subsection{Adaptive sparsity-aware load balancing}

\begin{figure}[thpb]
    \vspace{-3mm}
    \centering
    \includegraphics[width=\linewidth]{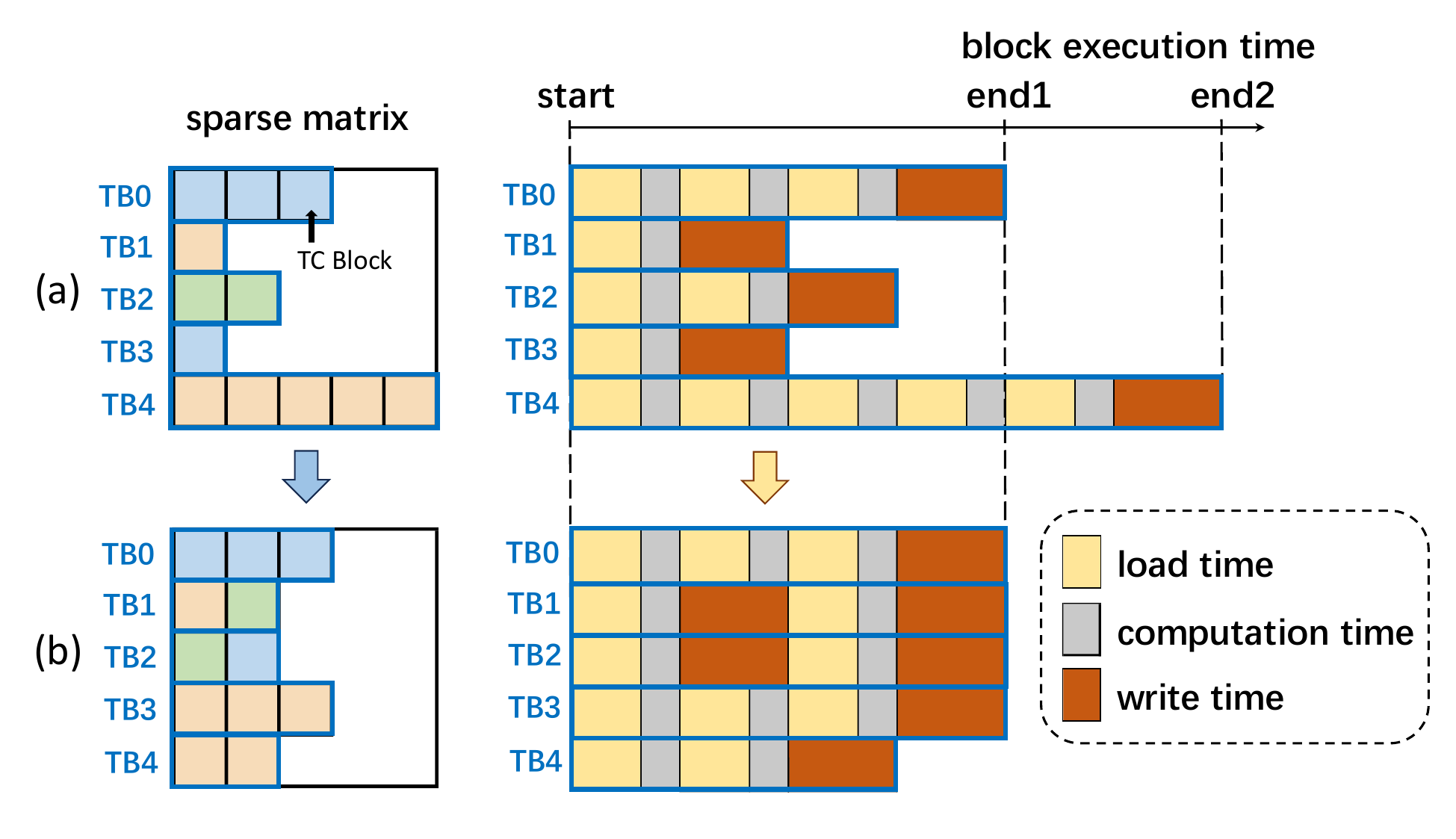}
    \caption{The diagram of adaptive sparsity-aware load balancing. (a) is without load balancing; (b) is with load balancing.}
    \label{Load Balance}
\end{figure}

Due to the various sparsity of different sparse matrices, the number of TC blocks in each RowWindow is imbalanced, which significantly impacts the efficiency of SpMM computations.
We propose an adaptive sparsity-aware load balancing method that dynamically rearranges TC blocks for each TB based on various sparsity characteristics, as shown in Figure \ref{Load Balance}.

Without load balancing, each TB processes all the TC blocks of a single RowWindow and writes the multiplication results back to global memory only once.
In contrast, with load balancing, cross-row write-back is introduced, as a TB may compute TC blocks from multiple RowWindows. Furthermore, concatenating RowWindows results in additional memory accesses for dense matrix $B$ and dense matrix $C$.
The load balancing method in DTC-SpMM is shown in Figure \ref{Load Balance} (a). There is one TC block for the second TB (TB1), resulting in one load time, one computation time, and one write time.
After considering the data write-back time for each TB and implementing a new load balancing approach, we design a method to rearrange the TC blocks for each TB, as shown in Figure \ref{Load Balance} (b). In this approach, the second TB is filled with two TC blocks from different RowWindows, and the fourth RowWindow in Figure \ref{Load Balance} (a) is split across two TBs, resulting in additional write-back overhead. Since load balancing introduces extra overhead, deciding whether to apply it is crucial. For some sparse matrices, whose computation is already well-balanced, no further adjustments are necessary. Only imbalanced matrices require adjustments to enhance performance.

We define a metric, $IBD$, as shown in Equation (\ref{equation-judge-load-balance}), to measure the degree of imbalance in the sparse matrix, where $TCblockPerRowWindow$ represents the number of TC blocks in each RowWindow, $AvgTCblock$ represents the average number of TC blocks across all RowWindows in the matrix, and $NumOfRowWindow$ represents the total number of RowWindows. We set a threshold for $IBD$. When $IBD$ exceeds 8, we consider the matrix to be highly imbalanced, thereby necessitating the application of a load balancing method.

\begin{displaymath}
\small
\label{equation-judge-load-balance}
\begin{split}
IBD = \frac{\sum\lvert TCBlockPerRowWindow - AvgTCBlock \rvert}{NumOfRowWindow}
\end{split}
\tag{3}
\end{displaymath}


Taking into account both hardware characteristics and matrix sparsity, we propose an adaptive sparsity-aware load balancing method that dynamically rearranges TC blocks between TBs based on pipeline evaluations. We develop a performance model, as shown in Equation (\ref{equation-perf-model}), which enables the reallocation of each TB's load, automatically assesses pipeline time, and optimizes load distribution.
In Equation (\ref{equation-perf-model}), $T$ represents the SpMM computation time for each TB using our proposed load balancing method, $LoadDenseTime$ represents the time required to load a dense matrix $B$ tile, $MMATime$ represents the time taken for the TC mma calculation, and $WBTime$ represents the time required to write back the computation result $C$. $FeatureDim$ represents the dimension of dense matrix B, $TcBlockPerTB$ represents the number of TC blocks in each TB, $Bandwidth$ is used to represent the theoretical memory bandwidth across different architectures, and $FLOPS$ represents the theoretical TF32 FLOPS of the TCU.
$M$ represents the number of rows of a sparse matrix $A$ tile, K represents the number of columns of a sparse matrix $A$ tile, and N represents the number of columns in a dense matrix $B$ tile. After swapping the left-hand matrix and the right-hand matrix, the dimensions $M$, $K$, and $N$ are set to 8, 8, and 16, respectively.

\begin{displaymath}
\label{equation-perf-model}
\small
\begin{split}
T & = LoadDenseTime + MMATime + WBTime \\
     & = \frac{K \times FeatureDim \times TcBlockPerTB}{Bandwidth} \\
     & + \frac{M \times (2 \times K - 1) \times FeatureDim}{FLOPS} \\
     & + \frac{K \times FeatureDim \times TcBlockPerTB}{Bandwidth}
\end{split}
\tag{4}
\end{displaymath}

Based on Equation (\ref{equation-perf-model}), the number of TC blocks assigned to each TB is redistributed to ensure nearly uniform computation time across all TBs.
We set a maximum threshold of 32 TC blocks per TB to facilitate this redistribution.

\section{Experiments}

\subsection{Experimental Setup}

We compare the overall performance of Acc-SpMM with state-of-the-art SpMM kernels on TCs and CUDA Cores, including TCGNN-SpMM \cite{wang2023tc}, DTC-SpMM \cite{Fan2024DTCSpMMBT}, SparseTIR \cite{Ye2022SparseTIRCA}, Sputnik \cite{Gale2020SparseGK}, and the widely used SpMM kernel in the cuSPARSE library \cite{Naumov2010}. All experimental results are obtained under the environment configured with CUDA version 11.8 or higher.
In addition to 10 representative real-world matrices listed in Table \ref{dataset}, a total of 414 matrices from the SuiteSparse Matrix Collection \cite{Davis2011TheUO} are evaluated, which are consistent with the matrices used in DTC-SpMM. Based on $AvgL$, we categorize the datasets into two types: type-1 matrices, which have a small $AvgL$, and type-2 matrices which have a large $AvgL$. All experiments are conducted on current mainstream NVIDIA GPU architectures, as detailed in Table \ref{gpu-archi}, including RTX 4090 (Ada Lovelace) \cite{nvidia_rtx_4090_architecture}, A800 (Ampere) 80GB PCIe \cite{nvidia_a100}, and H100 (Hopper) 80GB SXM \cite{H100_Tensor_Core_GPU_Architecture}.
\begin{table}[ht]
\small
\caption{Datasets for evaluation (Abbr: abbreviation, $AvgL$: average nnz length in each row of dataset). 1:from TC-GNN\cite{wang2023tc}; 2:from SNAP\cite{leskovec2016snap}; 3:from DGL\cite{wang2019deep}; 4:from OGB\cite{hu2020open}.}
\label{dataset}
\begin{tabular}{@{\extracolsep\fill}l|ccccc}
\hline
\  & Dataset     & Abbr.  & Row\&Col & NNZ & AvgL \\
\hline
1 & YeastH\textsuperscript{1}      & YH    & 3,138,114 & 6,487,230 & 2.07 \\
& OVCAR-8H\textsuperscript{1}    & OH    & 1,889,542 & 3,946,402 & 2.09 \\
& Yeast\textsuperscript{1}       & Yt    & 1,710,902 & 3,636,546 & 2.13 \\
& roadNet-CA\textsuperscript{2}    & rCA   & 1,971,281 & 5,533,214 & 2.81 \\
& roadNet-PA\textsuperscript{2}    & rPA   & 1,090,920 & 3,083,796 & 2.83 \\
& DD\textsuperscript{1}            & DD    & 334,926  & 1,686,092 & 5.03 \\
& web-BerkStan\textsuperscript{2}   & WB    & 685,230 & 7,600,595 & 11.09 \\
\hline
2 & FraudYelp-RSR\textsuperscript{3}     & FY-RSR    & 45,954 & 6,805,486 & 148.09 \\
& reddit\textsuperscript{3}        & reddit    & 232,965 & 114,848,857 & 492.99 \\
& protein\textsuperscript{4}       & protein    & 132,534 & 79,255,038 & 598.00 \\
\hline
\end{tabular}
\end{table}

\begin{table}[ht]
\small
\caption{The GPU architectures used for experiments (MEM: memory, TF32(TFLOPS): dense TC TF32 tera floating point operations per second, BW: bandwidth).}
\label{gpu-archi}
\begin{tabular}{@{\extracolsep\fill}c|ccc}
\hline
  GPU     & MEM Type  &\makecell{TF32(TFLOPS)}  &MEM BW\\
\hline
RTX 4090\cite{nvidia_rtx_4090_architecture}   & 24GB GDDR6X & 82.6  &   1008GB/s\\
A800\cite{nvidia_a100}       & 80GB HBM2  & 156  & 1935GB/s\\
H100\cite{H100_Tensor_Core_GPU_Architecture}       & 80GB HBM3   & 494.7  & 3.35TB/s\\
\hline
\end{tabular}
\end{table}

We measure the average performance with the dense matrix $B$ columns of 128, 256, and 512 using tensor float32 (TF32) precision. Additionally, we obtain a geomean result over the 414 matrices from the SuiteSparse Matrix Collection across different architectures, along with detailed results in GFLOPS for these matrices.

\subsection{Overall evaluation}

Using cuSPARSE on CUDA Cores as the baseline, our proposed Acc-SpMM achieves significant performance improvements over state-of-the-art SpMM kernels on both TCs and CUDA Cores, for both the 10 matrices listed in Table \ref{dataset} and the 414 matrices from the SuiteSparse Collection.

\begin{figure}[thpb]
    \centering
    \includegraphics[width=\linewidth]{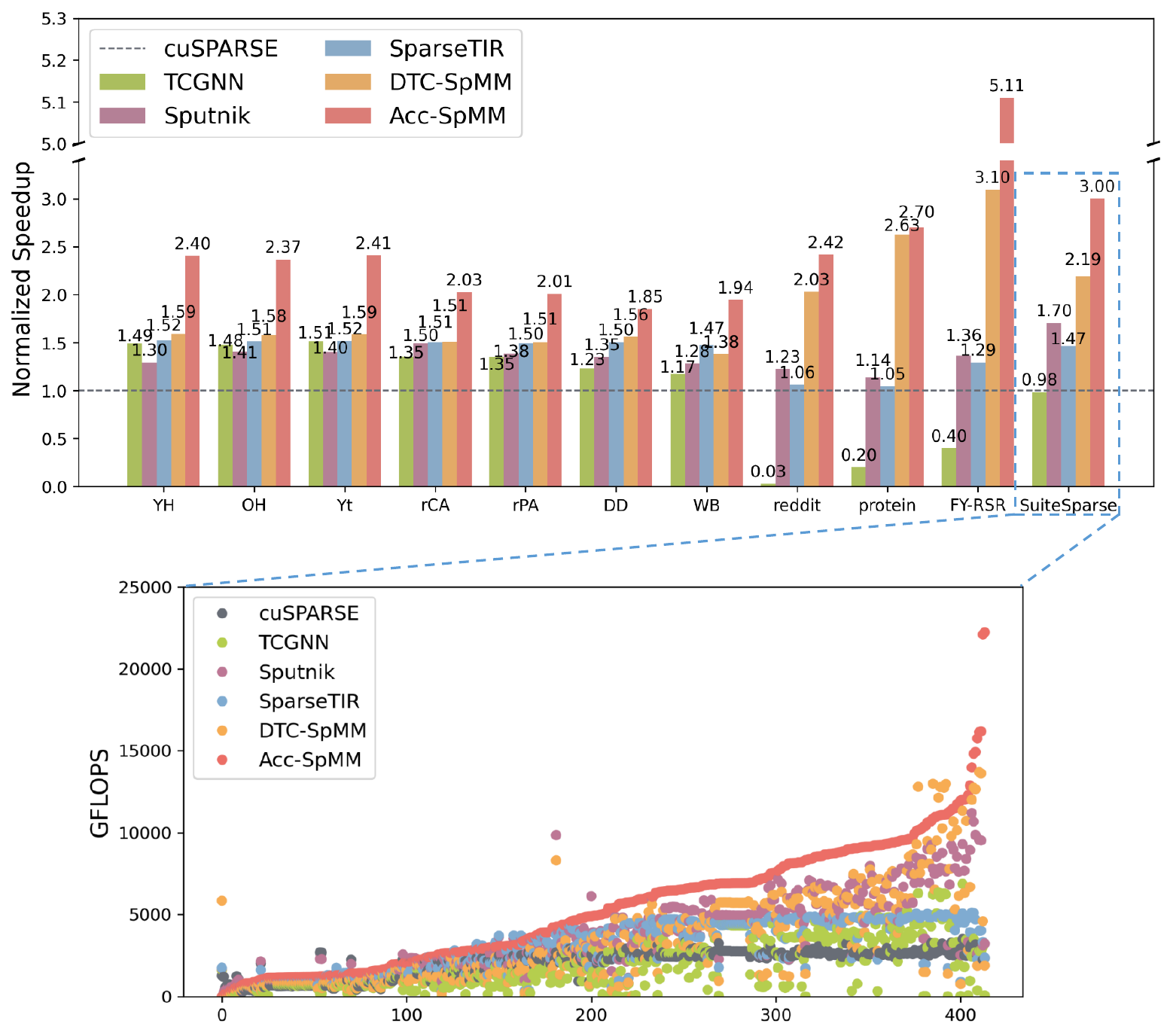}
    \caption{The performance of normalized speedup and detailed Gflops on RTX4090.}
    \label{RTX4090}
\end{figure}



\begin{figure}[thpb]
    \centering
    \includegraphics[width=\linewidth]{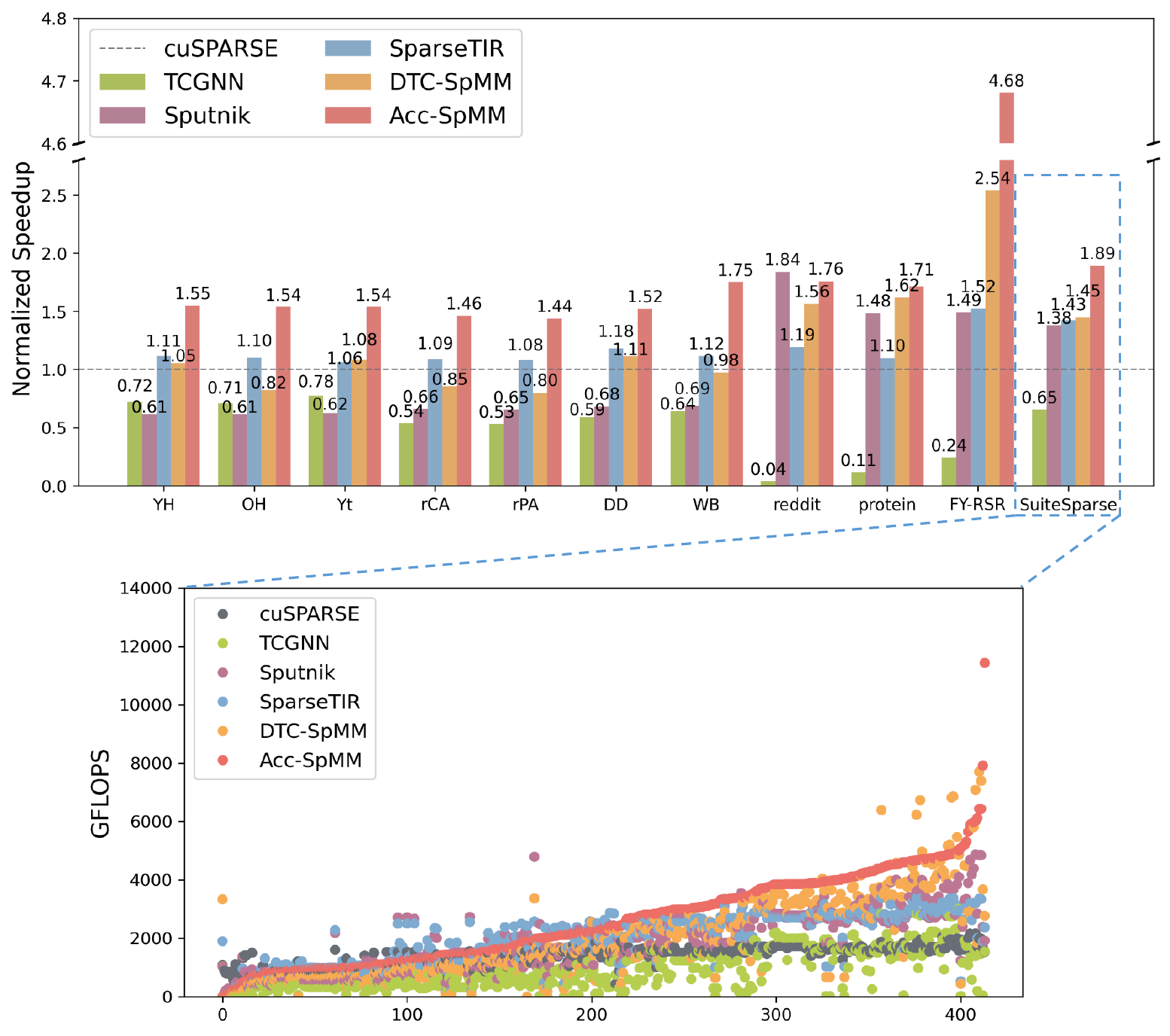}
    \caption{The performance of normalized speedup and detailed Gflops on A800.}
    \label{A800}
\end{figure}

\begin{figure}[thpb]
    \centering
    \includegraphics[width=\linewidth]{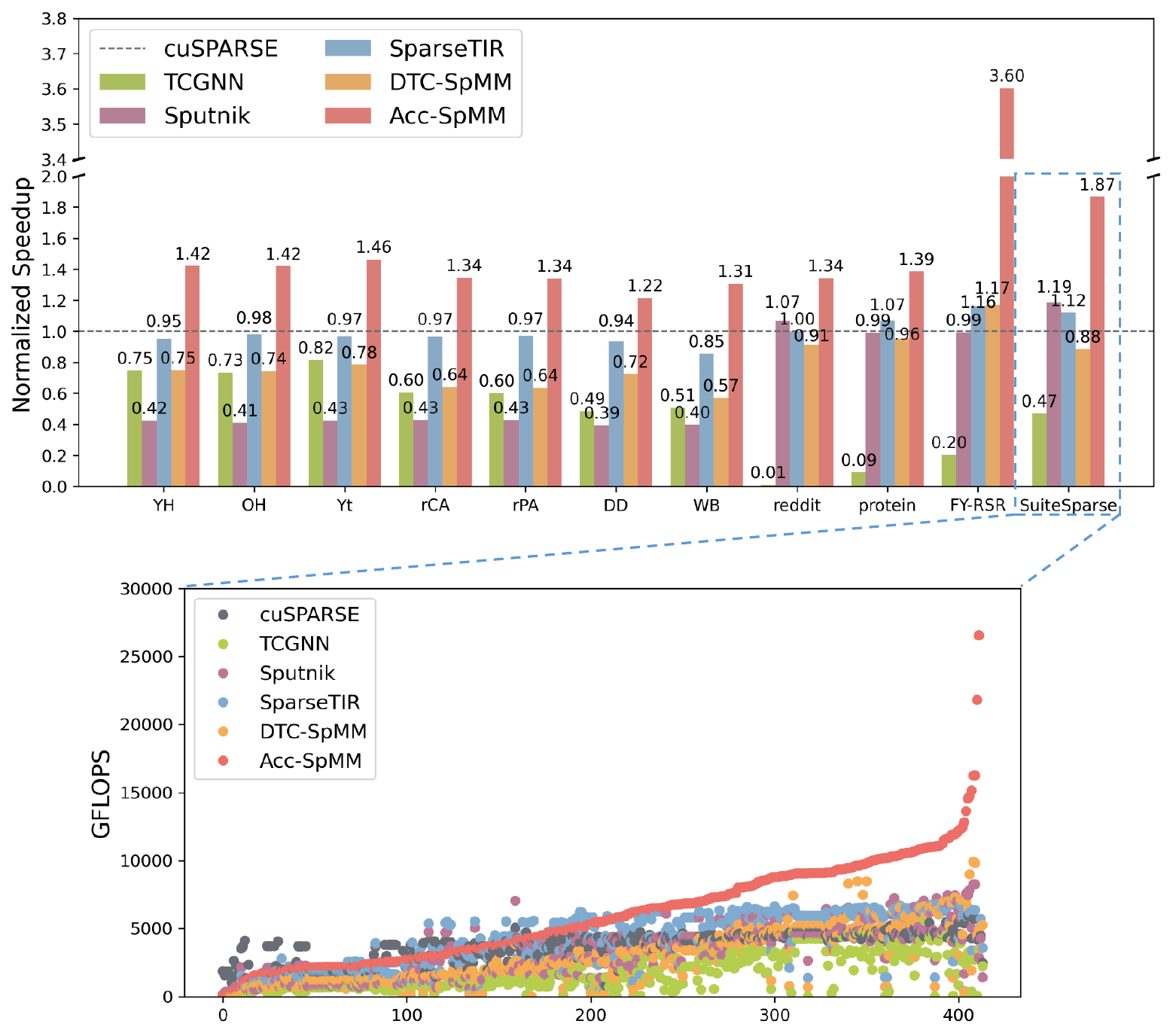}
    \caption{The performance of normalized speedup and detailed Gflops on H100.}
    \label{H100}
\end{figure}

On various NVIDIA GPU architectures, our proposed Acc-SpMM has always maintained good performance. Notably, on H100, where the performance of cuSPARSE-SpMM has been greatly improved, Acc-SpMM still achieves a significant speedup. All experiments are conducted on the GPUs listed in Table \ref{gpu-archi}, with cuSPARSE serving as the baseline, represented by the grey dashed line.
Figures \ref{RTX4090}, \ref{A800}, and \ref{H100} illustrate the performance comparisons on RTX 4090, A800, and H100, respectively. Acc-SpMM achieves significant speedups over the 10 representative real-world sparse matrices, with an average of 2.52× on RTX 4090, 1.91× on A800, and 1.58× on H100. The speedup is even more pronounced for type-2 matrices, reaching up to 5.11× on RTX 4090, 4.68× on A800, and 3.60× on H100. 

In Figure \ref{RTX4090}, 
Acc-SpMM outperforms other methods both on TCs and on CUDA Cores across all matrices. Compared to SpMM kernels CUDA Cores, Acc-SpMM leverages techniques such as efficient memory compression format, reordering, pipelining, and load balancing to fully utilize TCUs, which can perform more FMA operations per cycle compared to CUDA cores, enabling better performance to CUDA Core implementations, such as Sputnik, SparseTIR, and cuSPARSE. While TCGNN optimizes compressed storage formats to better utilize TCs, it still introduces significant redundancy. DTC-SpMM improves locality through reordering, along with pipelining and load balancing, but leaves opportunities for further optimization. Acc-SpMM effectively addresses these limitations, achieving further improvements in performance compared to both TCGNN and DTC-SpMM.
Figure \ref{A800} illustrates the the performance on A800. Notably, Sputnik demonstrates superior performance on the Reddit dataset due to its high number of nnzs. By effectively managing non-contiguous memory accesses, Sputnik significantly reduces memory access overhead. This optimization is especially beneficial on the A800, a streamlined version of the original architecture, leading to noticeable performance improvements.
The SpMM operator is a memory-bound operation. The H100 SXM leverages HBM3 memory, significantly boosting memory bandwidth and accelerating memory access. Additionally, the H100 hardware architecture is designed with optimizations for sparsity, allowing dynamic sparsity to be fully utilized. As a result, cuSPARSE shows a significant performance improvement on H100, as shown in Figure \ref{H100}. However, Acc-SpMM maintains superior performance due to optimizations in reordering, compressed formats, pipelining, and load balancing.

\subsection{Detailed evaluation}

\subsubsection{Evaluation of data-affinity based reordering.}

\begin{figure}[ht]
  \centering
  \begin{minipage}{0.67\columnwidth}
  \includegraphics[height=3.7cm]{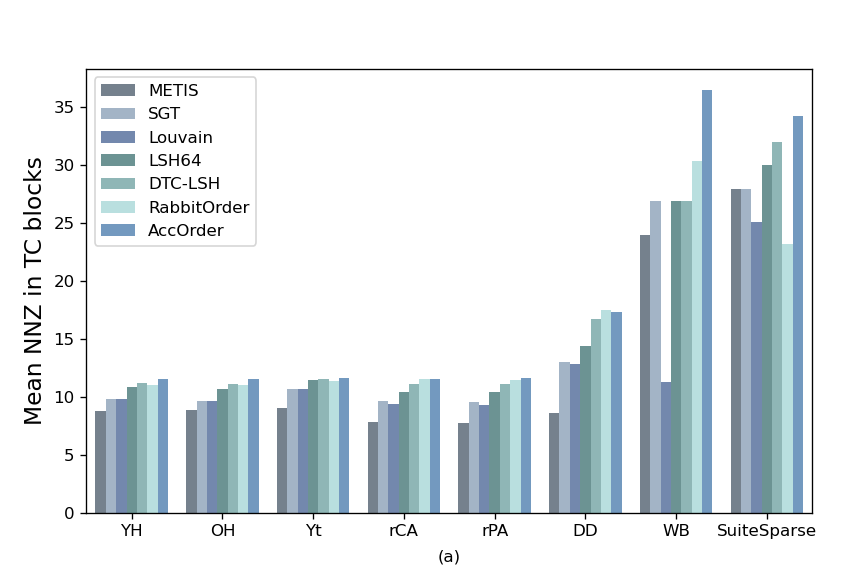} 
  \end{minipage}\hfill
  \begin{minipage}{0.33\columnwidth}
    \includegraphics[height=3.7cm]{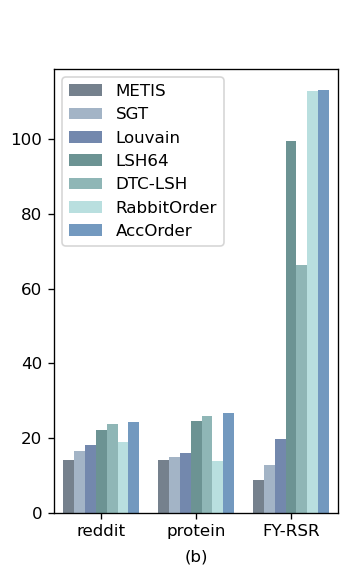} 
  \end{minipage}
  \caption{Comparisons on MeanNNZTC with different reordering method.}
  \label{meanNnzInTC}
\end{figure}

Compared to state-of-the-art reordering algorithms, our proposed reordering algorithm significantly improves data density within TC blocks, as shown in Figure \ref{meanNnzInTC}.
We compare our proposed reordering algorithm against six commonly used algorithms: METIS \cite{karypis1998software}, Louvain \cite{Que2015ScalableCD}, SGT \cite{wang2023tc}, LSH64 \cite{Huang2021UnderstandingAB}, DTC-LSH \cite{Fan2024DTCSpMMBT}, and Rabbit Order \cite{Arai2016RabbitOJ}. We define the MeanNNZTC as the average number of nnzs in each TC block, which serves as a metric to evaluate the effectiveness of reordering algorithms. Our proposed reordering algorithm achieves the highest MeanNNZTC blocks across all the evaluated sparse matrices.

\begin{figure}[ht]
  \centering
  \includegraphics[width=0.95\columnwidth,height=3cm]{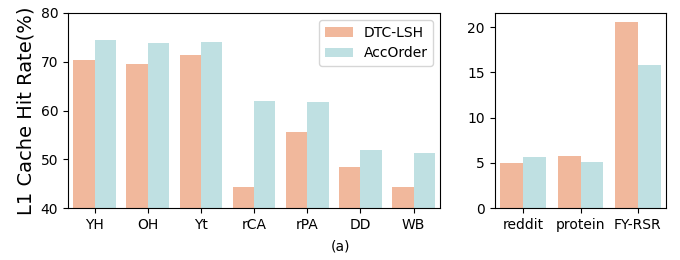}
  \includegraphics[width=0.95\columnwidth,height=3cm]{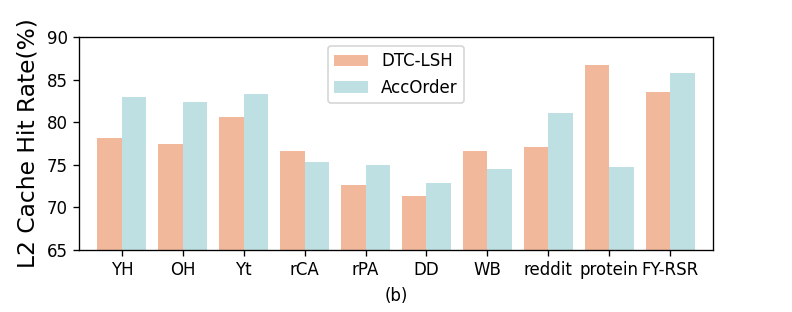}
  \caption{Experiments on A800 (a) L1 cache hit rate and (b) L2 cache hit rate.}
  \label{cachehitrate}
\end{figure}

Compared to DTC-LSH and Rabbit Order, our algorithm efficiently improves the MeanNNZTC across all 10 matrices as well as the 414 matrices from the SuiteSparse Collection, with average gains of 1.28× and 1.10×, respectively. Additionally, it outperforms the improvements achieved by METIS, SGT, Louvain, and LSH64.
As shown in Figure \ref{meanNnzInTC}, the improvement becomes more significant as the $AvgL$ increases.
Figure \ref{cachehitrate} (a) demonstrates significant improvements in L1 cache hit rate across most datasets, with a peak increase of 17.56\%.
Similarly, as depicted in Figure \ref{cachehitrate}(b), our algorithm significantly boosts L2 cache performance, also achieving a maximum increase of $4.93\%$.

It should be noted that in order to be consistent with DTC-SPMM, we only reorder the sparse matrix and do not perform corresponding row reordering on the dense matrix. If the dense matrix undergoes row reordering, the overall performance and cache hit rate will be further improved.

\subsubsection{Evaluation of memory efficient compressed format.}

\begin{figure}[thpb]
    \centering
    \includegraphics[width=\linewidth]{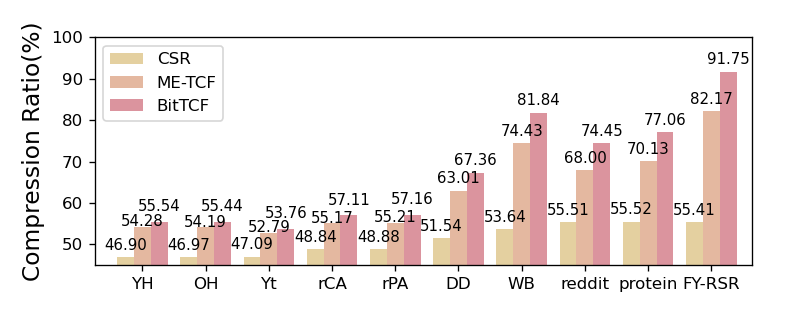}
    \caption{Comparisons of compression ratio among CSR, ME-TCF and BitTCF based on TCF \cite{wang2023tc}.}
    \label{compression ratio}
\end{figure}

Figure \ref{compression ratio} illustrates the comparison of compression ratios among TCF \cite{wang2023tc}, CSR format \cite{Saad2003}, ME-TCF \cite{Fan2024DTCSpMMBT}, and  BitTCF. Based on the memory usage of TCF, the compressed format used in TCGNN, BitTCF achieves the highest compression ratio, averaging 16.12\% higher than the CSR format and 4.21\% higher than ME-TCF. In addition to improving the compression ratio, BitTCF also reduces format conversion overhead, achieving a 15\% decrease compared to ME-TCF when converting from the CSR format.

The reduction in storage format overhead primarily depends on two factors: the use of the $TCLocalBit$ array to represent the positions of nnzs within each TC block, and the reordering of nnzs in the original matrix to make their arrangement more compact, which significantly reduces the number of TC blocks generated.
The TCF stores information about both zero elements and nnzs, whereas CSR only stores nnzs. ME-TCF further compresses the reordered matrix by using int8 to store the relative position of each nnz. Building on the reordered matrix, BitTCF utilizes a fixed uint64 to store the positions of nnzs within each TC block, achieving further compression of sparse matrix storage compared to ME-TCF. 
We choose an $8 \times 8$ tile as the block size for matrix A, which also conveniently allows the use of uint64 to encode the positions of nnzs.
Moreover, it only takes few seconds to covert from CSR to BitTCF. For iterative applications, the overhead of this conversion is minimal and can be amortized.
\subsubsection{Evaluation of high-throughput pipeline.}

\begin{figure}[thpb]
    \centering
    \includegraphics[width=\linewidth]{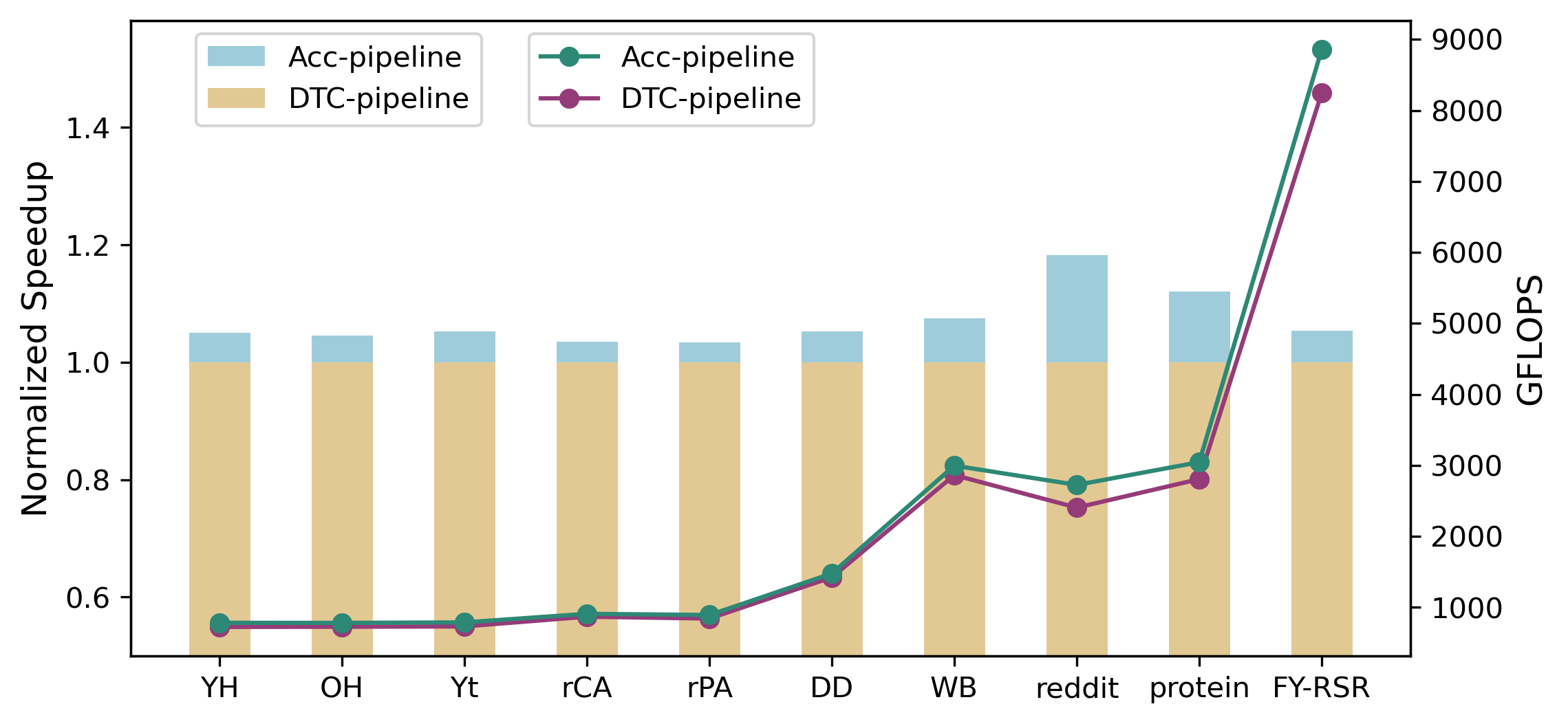}
    \caption{Comparisons of speedup and GFLOPS between DTC-pipeline and Acc-pipeline on A800.}
    \label{pipeline}
\end{figure}

In Figure \ref{pipeline}, the performance and speedup comparisons between our proposed pipeline and the DTC-pipeline are illustrated.
The purple line represents the Gflops of the DTC-pipeline, while the green line represents the Gflops of the Acc-pipeline.
Compared to the DTC-pipeline, which is specifically optimized for TCs, our proposed pipeline achieves higher performance. Using the DTC-pipeline as baseline, our proposed pipeline demonstrates speedups across all 10 sparse matrices. 
The improvement becomes more pronounced when the AvgL is larger, with average speedups of 1.06x and 1.16x for type-1 and type-2 matrices, respectively. Since a TB processes more TC blocks in type-2 matrices, it achieves a greater reduction in pipeline bubbles compared to type-1 matrices. As a result, the pipeline exhibits better performance on type-2 matrices than on type-1 matrices.

\subsubsection{Evaluation of adaptive load-balancing.}

\begin{figure}[thpb]
  \centering
  \includegraphics[height=3.1cm]{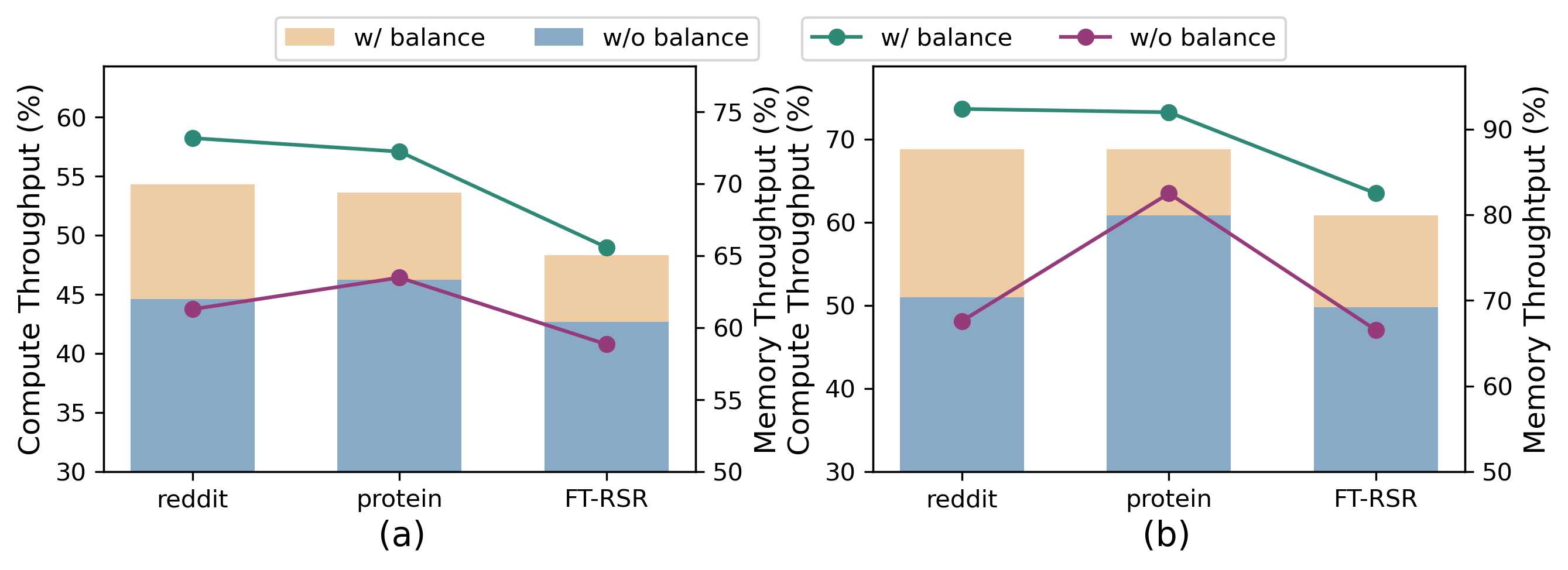} 
  \caption{Comparisons of compute throughput and memory throughput between with load balance and without load balance on A800 (a) and H100 (b).}
  \label{exp-load-balance}
\end{figure}

Figure \ref{exp-load-balance} shows the comparisons of throughput with and without load balancing on imbalanced matrices.
For type-1 matrices, each TB typically handles only one or two TC blocks, resulting in relatively balanced workloads across all TBs, as indicated by Equation (\ref{equation-judge-load-balance}). Therefore, load balancing is not necessary for type-1 matrices,
and we focus our load balancing experiments mainly on type-2 matrices.
The blue rectangles indicate compute throughput without load balancing, and the orange rectangles represent compute throughput with our proposed load balancing. The purple line shows memory throughput without load balancing, while the green line shows memory throughput with our proposed load balancing. From  Figure \ref{exp-load-balance}, it can be concluded that our proposed load-balancing effectively improves both compute throughput and memory throughput.

\subsubsection{The ablation experiments.}

\begin{figure}[thpb]
    \centering
    \includegraphics[width=\linewidth]{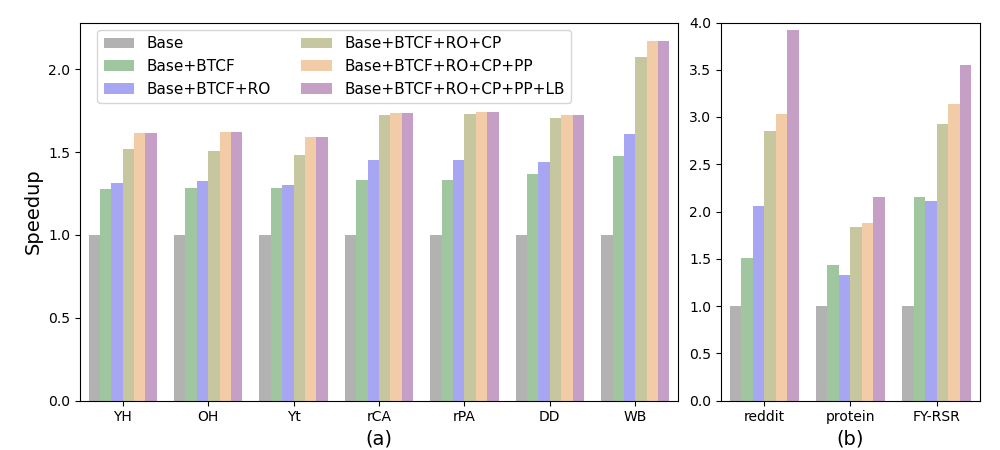}
    \caption{The ablation study on H100 with a feature dim of 128. Base: DTC-SpMM method without load balancing; BTCF: BitTCF; RO: Data-affinity reordering; CP: Cache policy control; PP: Pipeline; LB: Load balancing.}
    \label{ablation}
\end{figure}

To evaluate the effectiveness of different optimization techniques on the overall performance of Acc-SpMM, we take DTC-SpMM without load balancing as the baseline and conduct ablation experiments on H100 with dense matrix $B$ columns of 128, as shown in Figure \ref{ablation}. The BitTCF not only enhances storage efficiency through compression but also boosts computational performance by minimizing the amount of data transferred from global memory to registers. 
The data-affinity-based reordering increases MeanNNZTC, thereby reducing the number of TC blocks in the sparse matrix and effectively improving performance. However, for the protein dataset, our reordering method improves the MeanNNZTC but reduces both L1 and L2 cache hit rates, as shown in Figure \ref{cachehitrate}. Similarly, for the FY-RSR dataset, our method decreases the L1 cache hit rate. As a result, the performance of these two datasets slightly declines after applying the reordering method. The cache policy effectively utilizes L1 and L2 caches, improving cache hit rates and consequently boosting performance. The pipeline leverages the GPU’s high memory access bandwidth to enable multiple overlaps between mma computation and various memory accesses, including fetching matrix $A$ tiles and indices. For type-1 matrices, the performance improvement from pipelining is less pronounced compared to type-2 matrices. This is mainly because, in type-2 matrices, each TB is assigned more TC blocks for computation, resulting in a greater reduction of pipeline bubbles during iterations and leading to more significant performance gains. The load balancing method ensures more balanced computation times across TBs and innovatively incorporates write-back time into the pipeline performance model, significantly enhancing the model’s accuracy and improving load balancing effectiveness.

\section{Related Work}

Given an m-by-k sparse matrix $A$ and a k-by-n dense matrix $B$, SpMM computes $A$ multiply $B$ and obtains an m-by-n dense matrix $C$. 
With the architectural advancements and technique updates in GPUs, 
there are various methods to optimize SpMM on GPUs, targeting both CUDA cores and TCs. 
The widely-used NVIDIA cuSPARSE library \cite{Naumov2010} implements several basic linear algebra subroutines for sparse matrices
, offering high-performance SpMM kernels for both CUDA cores and Tensor Cores.
Hone et al. proposed RS-SpMM \cite{Hong2018EfficientSM} and ASpT \cite{Hong2019AdaptiveST} 
, which use adaptive tiling to partition sparse matrices into dense and sparse blocks, enabling the reuse of the dense matrix in shared memory.
Jiang et al. proposed a row-reordering method \cite{Jiang2020AND} based on ASpT, which obtains more dense matrix blocks for a sparse matrix. 
Huang et al. proposed GE-SpMM \cite{Huang2020GESpMMGS} to improve the efficiency of global data access by reusing the sparse matrix. 
Gale et al. proposed Sputnik \cite{Gale2020SparseGK}, which introduces a one-dimensional tiling scheme and reverse offset memory alignment to improve the memory access efficiency of sparse matrices.
Pang et al. proposed row decomposition method RoDe \cite{Pang2024ARD}, which splits the sparse matrix into load-balanced blocks and utilizes sub-block pipeline to improve memory access efficiency. 

In recent years, there has been increasing attention on accelerating SpMM on TCs.
cuSPARSELt \cite{NVIDIA_2020} is a library that supports Sparse Tensor Cores and imposes strict constraints on structured sparsity, specifically following a 2:4 pattern.
Chen et al. proposed VectorSparse \cite{Chen2021EfficientTC}, an evolution of Sputnik, which leverages one-dimensional groups of sparse elements to enhance data locality.
Castro et al. proposed CLASP \cite{Castro2022ProbingTE} with a new column-vector pruning-aware implementation of SpMM to improve data locality. 
Chen et al. proposed tensor-core-based one-dimensional octet tiling method \cite{Chen2021EfficientTC} to improve memory access efficiency. 
Li et al. proposed Magicube \cite{Li2022EfficientQS} with a novel sparse compression format SR-BCRS to accelerate SpMM using low-precision integers.
Wang et al. proposed TC-GNN \cite{wang2023tc}, the first GNN acceleration framework on TCs for GPUs, featuring an optimized general SpMM kernel that utilizes graph sparsity patterns to tile the sparse matrix through effective structural manipulation.

\section{Conclusion and Future Work}

In this paper, we propose Acc-SpMM, a high-performance SpMM library on TCs based on systematic optimizations. These optimizations involve high-level algorithms, low-level memory access optimizations, and the instruction pipeline as well. 
Through performance evaluation over 10 representative large-scale power-law graph matrices in GNNs and 414 sparse matrices from Suite Sparse Collection \cite{Davis2011TheUO} on recent mainstream NVIDIA GPU architectures, Acc-SpMM achieves an average speedup of 2.52x (with a maximum of 5.11x) on the RTX 4090, 1.91x on average (up to 4.68x) on the A800, and 1.58x on average (peaking at 3.60x) on the H100 (SXM) when compared to cuSPARSE. 

In the future, we plan to reorder the columns of the sparse matrix while simultaneously reordering the rows of the dense matrix, further improving cache hit rates. We will also optimize the implementation of reordering algorithm to reduce overhead. Additionally, we will integrate the SpMM operator into DGL for practical use in GNNs, making our approach more systematic and improving the feasibility in end-to-end applications. 




\section*{Acknowledgments}

We thank all the anonymous reviewers for their constructive comments and suggestions, which greatly improved the quality of this work. This work was supported by the National Key Research and Development Program of China (2023YFB3002100) and the Chinese Academy of Sciences Strategic Priority Research Program (XDB0500103). 

\bibliography{ref}

\clearpage
\newpage
\pagestyle{empty}

\end{document}